\documentstyle[aps]{revtex}
\begin{document}
\begin{title}
{\bf Negative  Electron-electron Drag \\Between Narrow Quantum Hall Channels}
\end{title}
\author{H. C. W. Tso$^{*}$, D. J. W. Geldart$^{*}$, and P. 
Vasilopoulos$^{\dagger}$}
\address{$^{*}$ Dalhousie University, 
Department of Physics,  Halifax, Nova
Scotia, Canada B3H 3J5
 \\
$^{\dagger}$Concordia University, Department of Physics,  1455 de
Maisonneuve Blvd. West, \\Montr\'{e}al,
Qu\'{e}bec, Canada H3G 1M8}
\date{October 9, 1997}
\maketitle
\begin{abstract}
Momentum transfer due to Coulomb interaction between two parallel, 
two-dimensional, {\it narrow}, and spatially separated layers, when a 
current $I_{\rm drive}$ is driven through one layer, is studied in the 
presence of a perpendicular magnetic field $B$.  The current induced 
in the drag layer, $I_{\rm drag}$, is evaluated self-consistently with 
$I_{\rm drive}$ as a parameter.  $I_{\rm drag}$ can be positive or 
negative depending on the value of the filling factor $\nu$ of the 
highest occupied bulk Landau level (LL).  For a fully occupied LL, 
$I_{\rm drag}$ is {\it negative}, (i.e., it flows opposite to $I_{\rm 
drive}$), whereas it is positive for a half-filled LL. When the 
circuit is opened in the drag layer, a voltage $\Delta V_{\rm drag}$ 
develops in it; it is negative for a half-filled LL and positive for a 
fully occupied LL. This {\it positive} $\Delta V_{\rm drag}$, 
expressing a {\it negative} Coulomb drag, results from energetically 
favored {\it near-edge inter-LL transitions} that occur when the 
highest occupied bulk LL and the LL just above it become degenerate.
\end{abstract}
\maketitle
\section{INTRODUCTION}

Recently the transresistance $R_T$ between two parallel 
two-dimensional (2D) systems~\cite{tsoetal,others,vm,fin} has been 
studied extensively due to advances in measuring~\cite{exp} 
techniques.  In most previous studies~\cite{tsoetal,others,vm,fin,exp} 
only wide systems were considered, in which edge effects can be 
ignored.  However, edge effects become prominent in narrow systems 
especially when a magnetic field $B$ is present and the highest bulk 
LL is completely occupied.  The electronic edge states are very 
different from the bulk states.  For instance, if a weak random 
impurity potential is present, the bulk states are 
disordered~\cite{halperin} and occur in a series of energy bands of 
finite width $\Gamma_n (k)$, centered about the LLs with no states in 
the region between them.  On the other hand, the edge states are 
degenerate with respect to the LLs.  This can lead to 
profound differences in transport properties between the case when the 
highest bulk LL is completely occupied and that when it is not.

In this paper we study the influence of edge states on the drag by 
solving {\it self-consistently} the Coulomb-coupled Schr\"{o}dinger 
equations for two {\it narrow} Hall bars.  This goes substantially 
beyond the simple, not self-consistent treatments of Ref.  \cite{efr} 
that used a parabolic confining potential.  Here the potential is 
evaluated self-consistently from an initial square-well potential.  We 
find that a current $I_i$ is always induced in the direction of 
$I_{\rm drive}$ and the Coulomb interaction lifts the degeneracy of 
all occupied edge LLs, but not that of unoccupied and 
neighbouring-occupied edge LLs.  The Coulomb drag obtained is {\it 
two-to-three orders of magnitude larger} than that in zero field due 
to the increase in the available density of states when an external 
magnetic field is present.  As the filling factor $\nu$ of the highest 
LL approaches 1, {\it near-edge inter-LL transitions} occur between 
this LL and that just above it.  These transitions make $I_{\rm drag}$ 
or $\Delta V_{\rm drag}$ change sign when the circuit in the drag 
layer is closed or open, respectively.  Thus, this negative drag is 
neither due to a thermal gradient~\cite{bor} nor to to an 
electron-hole coupling.

 In the next section we present the formalism.  In Sec.  III we  present and 
discuss 
 the numerical results.  We conclude with remarks  in Sec.  IV .
 
\section{FORMALISM}
\subsection{ Coupled Schr\"{o}dinger equations}

Consider two quantum Hall bars parallel to the $(x,\ y)$ plane, 
separated by a distance $d$ along the $z$ axis, of thickness zero, 
width $L_x=w$, and length $L_y\equiv L \gg \ell_c$.  The electrons are 
confined along $x$ by infinitely high potential barriers.  In a field 
${\bf B} = -B\ {\bf \hat z}$, the electron wavefunction in the drive 
layer $a$, when the circuit in the $y$ direction is closed, 
has the 
form $\psi_{nk}^a (x)\ e^{iky}/\sqrt{L}$ and $\psi^a$ obeys

\begin{equation}
-{\hbar^2\over{2m^*}}[{{\partial^2}\over{\partial x^2}} + 
 ( k-x/\ell_c^2)^2 - {e^*}^2\phi_{ab} (x)]\psi_{nk}^a 
(x) = E_{nk}^a \psi_{nk}^a (x).  \label{1.1}
\end{equation}
The corresponding wavefunction $\psi_{nk}^b (x)$ in the drag layer $b$ obeys

\begin{equation}
-{\hbar^2\over{2m^*}}[{{\partial^2}\over{\partial x^2}} + 
 ( k-x/\ell_c^2)^2 - {e^*}^2\phi_{ba} (x)]\psi_{nk}^b 
(x) = E_{nk}^b \psi_{nk}^b (x).  \label{1.11}
\end{equation}
Here $n$ is the LL number, $m^*$ the effective mass, 
$\ell_c=\sqrt{\hbar/eB}$ the magnetic length, $e^* ={2m^*\over\hbar^2} 
e/\sqrt{\epsilon}$, and $\epsilon$ the dielectric constant.  For 
simplicity spin is neglected.  The Coulomb potential $\phi_{ab}$
is given by 

\begin{equation}
\phi_{ab} (x) = -2 \int  dx^\prime\  [\delta\rho_{a} (x^\prime)
              \ln |x - x^\prime|
             +\delta\rho_{b} (x^\prime)
      \ln \sqrt{(x - x^\prime)^2 + d^2}], \label{1.2}
\end{equation}
As for the 
charge densities $\rho_a$ and $\rho_b$, they are given  by

\begin{equation}
\rho_a (x)= {1\over{2\pi}}\sum_n \int_{-k_0}^{k_0} dk\
         g_n^a (k) |\psi_{nk}^a (x)|^2  \ f(E_{nk+\delta k}^a).
\label{1.3}
\end{equation}
in the drive layer~\cite{wexler} and by

\begin{equation}
\rho_b (x)= {1\over{2\pi}}\sum_n \int_{-k_0}^{k_0} dk\
         g_n^b (k) |\psi_{nk}^b (x)|^2  \ f(E_{nk}^b).
\label{1.31}
\end{equation}
in the drag layer. In Eqs. (3)-(5)  $\delta\rho_\alpha (x) = 
\rho_\alpha (x) - \rho_{0 \alpha}$, $\alpha=a$ or $b$, $\rho_{0\alpha}$ is the 
background charge density,  $f$ is the Fermi function, and
$k_0 = w/2\ell_c^2$. Notice that  $\delta k=0$ in Eq. ($\ref{1.31}$) since  no
  current flows through in the drag layer. We take the effective 
background densities equal $\rho_{0a} = \rho_{0b}$ and constant.
 The Schr\"{o}dinger equations for the drive and drag layers are solved 
self-consistently.

The weight function $g_n^a (k)$ depends on $E_{nk}^a$ and expresses 
the degeneracy of the LLs.  When the $n$-th LL is completely filled, 
we have $g_n^a (k) = 1$.  When it is partially filled, we determine 
$g_n^a (k)$ self-consistently~\cite{tsos} from $E_{nk}^a$ and the 
level broadening $\Gamma_n^a (k)$ due to scatterers from the 
assumptions (i) $\Gamma_n^a (k)$ is independent of $k$, and (ii) the 
total neutrality of the Hall bar and the local neutrality at its 
center are maintained.  The meaning of these assumptions becomes clear 
if we consider two limiting cases for the average filling factor of 
the highest LL ${\bar\nu}\neq 1$.  First, for an infinitely wide bar 
without edges, the energy levels $E_{nk}^\alpha$ is definitely 
degenerate with respect to the wave vector $k$  
and everywhere we have the same filling factor $\nu=\bar \nu$.  Secondly, 
for a {\it narrow} Hall bar without
the $k$ degeneracy in $E_{nk}^\alpha$ and $g_n^\alpha (k) = 1$,
all  $k$ states are occupied if ${\bar\nu} = 1$.
But if ${\bar\nu} < 1$, not all $k$ states are occupied and electrons 
prefer those low-lying energy states near the center of the bar.  
Therefore the actual width of the Hall bar, defined by the density of 
electrons, will shrink to roughly ${\bar\nu} W$ and in it the filling factor is 
still $1$ 
even though $\bar\nu$ is less than 1.  In reality, there is scattering which
broadens $E_{nk}^\alpha$ by, say, $\Gamma_n^\alpha (k)$. Within
$\Gamma_n^\alpha (k)$, we assume that each $k$ has the same electron
occupancy, so the 
local filling factor depends on the local density of states at $k$.
This average electron occupancy depends on the density of states 
$N_n (k)$ and is reflected in 
\begin{eqnarray}
\nonumber
g_n (k)
=&& {\bar\nu}_n
{{N_n(0)}\over{N_n (k)}} \left[1-\Theta ({\bar\nu}_n
  N_n (0) - N_n (k))\right]\\*
  &&+ \Theta ({\bar\nu}_n N_n (0) - N_n (k)) \label{degen}
\end{eqnarray}
where $\Theta (x)$ is the Heaviside step function and ${\bar\nu}_n$ is 
the filling factor of the $n$th LL.  Consequently, the electron
density is nonzero within the width of the bar with local filling
factor $\nu \neq 1$.
At the center of the bar, which corresponds roughly to $k=0$, 
we have $g_n (k) = {\bar\nu}_n$.  This ensures the neutrality at the 
center which is in contrast with the single-layer treatment of Ref.  
\cite{wexler} where only the total neutrality was preserved and only 
the $\nu=1$ situation could be handled.  In our case, this local 
neutrality at the center ensures that $E_{nk} = (n + 
1/2)\hbar\omega_c$ at $k=0$.  In Eq.  ($\ref{1.3}$), $\delta 
k\parallel y$ is the shift of the Fermi surface~\cite{wexler} that 
results from the application of a current $I_{\rm drive}\parallel y$ 
through the drive layer in the presence of scattering.

\subsection{ Currents}

The drive and induced currents, $I_{\rm drive}$ and $I_i$, are
given by

\begin{equation}
I_{\rm drive}= -I_0 w^2\sum_n \int dx
           \int_{-k_0}^{k_0} dk\ g_n^a (k)
           (k - x/ {\ell_c^2}) |\psi_{n,k}^a (x)|^2
           f(E_{n,k+\delta k}^a),\label{1.4}
\end{equation}
and

\begin{equation}
I_{\rm i} = -I_0 w^2\sum_n \int dx
           \int_{-k_0}^{k_0} dk\ g_n^b (k)\
           (k - x/{\ell_c^2}) |\psi_{nk}^b (x)|^2 \
           f(E_{nk}^b).\label{1.5}
\end{equation}
where $I_0=e\hbar/2\pi m^* w^2$.  $I_{\rm i}$ consists not only of the 
current induced by momentum transfer but also of the classical ${\bf 
E}\times{\bf B}$-drift~\cite{plasma} current in the direction of ${\bf 
B }\times <{\bf E}>$ where $<{\bf E}>$ is the average of a finite 
electrostatic field ${\bf E}$ exerted on the drag electrons by the 
charges which accumulate at the edges of the drive layer and produce 
the Hall voltage.  This field ${\bf E}$ opens up the circular orbit of 
the drag electrons by accelerating those moving opposite to it and 
decelerating those moving in its direction~\cite{plasma}.  Thus it 
induces an average current $I_{\rm es}$ in the direction of $I_{\rm 
drive}$ parallel to ${\bf B} \times <{\bf E}>$ given by

\begin{equation}
I_{\rm es} = -I_0 w^2 \sum_n\int dx
            \int_{-k_0}^{k_0} dk\  g_n^{0} (k)\
           (k - x/\ell_c^2) |\chi_{nk} (x)|^2 \
           f(E_{nk}^{0}) ,\label{1.6}
\end{equation}
and $\chi_{nk} (x)$ obeys Eq.  (1) with $a\to b$, $\phi\to\bar\phi$, 
$\rho\to\bar\rho$, and $E\to\bar E$.  $\bar\phi (x) $ is given by Eq.  
(2) with the changes $\rho_a\to\bar\rho_b$, $\rho_b\to \rho_a$, and

\begin{equation}
\bar\rho_b (x)= {1\over{2\pi}}\sum_n \int_{-k_0}^{k_0} dk\
         g_n^{0} (k) |\chi_{nk} (x)|^2  \ f(E_{nk}^{0}).
\label{1.9}
\end{equation}
The use of the equilibrium energy $E_{nk}^{0}$ in the drag layer to 
derive $g_n^{0} (k)$ and of the Fermi function in Eq.  ($\ref{1.6}$) 
eliminates electron transitions between different $k$'s.  Thus the 
current due to momentum transfer is

\begin{equation}
I_{\rm drag} = I_i - I_{\rm es}.
\label{1.91}
\end{equation}
For fully occupied LLs we have $g_n^{0} (k) = g_n^b (k) = 1$ 
and  $I_{\rm drag}$ vanishes but the classical $<{\bf E}> \times {\bf B}$-drift 
$I_{\rm es}$ does not.

\subsection{ Explicit relation between the drag current and drag 
voltage.}

When the circuit in the drag layer is opened, apart from the voltage induced
by the $<{\bf E}> \times {\bf B}$-drift, a drag voltage $\Delta V_{\rm
drag} = LI_{\rm drag}\Delta t \delta(z)/\epsilon W$ develops along the 
bar as shown in Fig. 1. In a short time $\Delta t$ it produces an intermediate 
current,

\begin{equation}
I_{\rm int}
= {{\rho_{0} e\Delta V_{\rm drag}}\over B}
= \Delta V_{\rm drag} {{(n_{\rm max} + \nu)e^2}\over {2\pi\hbar}}
\end{equation}
 {\it across the width}, due to the
${\bf E_{\rm drag}} \times {\bf B}$-drift.
This results in an electric field $E_{\rm int} = I_{\rm int} \delta (z)
\delta t/\epsilon L$, across the width, which produces
 another ${\bf E}_{\rm int}\times{\bf B}$-drift
current {\it opposite} to $I_{\rm drag}$. The steady-state is reached when
these two currents are equal so that all the ${\bf E} \times {\bf
B}$-drifts are balanced. Thus, $E_{\rm int}$ is given by

\begin{equation}
E_{\rm int} = {{e^2 (n_{\rm max} + \nu)}\over{2\pi\hbar\epsilon L}}
  \Delta V_{\rm drag} \Delta t\ \delta (z)
  = {{2\pi\hbar}\over{e^2 w(n_{\rm max} + \nu)}} I_{\rm drag}
\end{equation}
with $n_{\rm max}$ the highest occcupied bulk LL number. The drag voltage is 
then

\begin{equation}
\Delta V_{\rm drag}
=-{{2\pi\hbar}\over{(n_{\rm max} + \nu)e^2}} {L\over w} I_{\rm
drag},
\end{equation}
where the minus sign is inserted according to the usual Ohm's law
convention; $I_{\rm drag}$ is given by Eq. (9).

\section{RESULTS AND DISCUSSION}

We consider a GaAs sample at zero temperature, with $d = 5 \ell_c$, $w 
= 20 \ell_c$, and $B$ chosen such that $\ell_c = 100 \AA$.    The 
Schr\"{o}dinger equations for the drive and drag layers are solved 
self-consistently and the average electron and current densities of 
the drag-layer, with $n_{\rm max} = 1$ and $\nu = 0.5$, are plotted in 
Fig.  2. The solid and dotted curves represent equilibrium and 
nonequilibrium quantities, respectively. Fig. 3 shows that the 
current density of the main figure is not 
symmetric. $E_{nk}^0$ and $g_n^0 (k)$ are shown in Fig. 4. 
The rapid change of $g_n^0 (k)$ near the edges reflects the structure of the 
boundaries between edge and disordered bulk states.  
As shown in Fig.  2, the current-density ($j_b$) plot has qualitatively five 
different 
zones, namely the outer-edge ($10 \ell_c > |x| > 6 \ell_c$) and 
inner-edge ($6 \ell_c > |x| > 3.5 \ell_c$) zones on both sides of the 
bar, and one bulk zone ($3.5 \ell_c > |x| > 0$).  In the outer-edge 
zone $j_b$ is opposite to that in the inner-edge zone and in the bulk 
zone it is negligible.  

To understand the results of Fig. 2 we consider the 
single-electron motion.  We first treat the equilibrium case ($I_{\rm 
drive} = 0$, $\delta k = 0$).  For finite ${\bf B}$, $j_b$ near the 
edge is affected by the sharp changes of the electron density $n(x)$ 
and of the effective confining potential.  Near the left edge, the 
gradient ${\bf\nabla} n(x)$ is so large that there are more electrons, 
with guiding center at $x+\delta x$, e.g., near the maximum of $n(x)$ 
at $x < -9\ell_c$, flowing into the paper ($y$ direction) than 
adjacent electrons, with guiding center at $x-\delta x$, flowing out 
of the paper; this gives rise to a local $j_b$, at $x$, in the 
negative $y$ direction.  This diamagnetic-drift current is thus 
proportional~\cite{plasma} to ${\bf\nabla} n(x)$.  In the same 
region the local effective electric field ${\bf E}_{\rm eff} 
=-{\bf\nabla} V_{\rm eff}$, produced by electrons in the center of the 
bar, pushes the electrons towards the edge.  This produces a ${\bf 
E}_{\rm eff} \times {\bf B}$-drift current~\cite{plasma} pointing in 
the same direction as the diamagnetic-drift current.  However, for 
electrons located further away from the edges ${\bf E}_{\rm eff}$ 
changes direction.  It pushes the electrons away from the edge towards 
the center of the bar, and produces a ${\bf E}_{\rm eff} \times {\bf 
B}$-drift $j_b$, proportional~\cite{plasma} to ${\bf\nabla} V_{\rm 
eff} (x) = - {\bf E}_{\rm eff}$ that points in the {\it positive} 
(negative) $y$ direction at the left (right) edge.  In most of the 
outer-edge region, the ${\bf E_{\rm eff}} \times {\bf B}$-drift is 
comparable to the diamagnetic-drift whereas in the inner-edge region 
the ${\bf E_{\rm eff}} \times {\bf B}$-drift dominates.  As for the 
bulk zone, ${\bf\nabla} n(x)$ and ${\bf\nabla} V_{\rm eff} (x)$ are 
almost zero and so is $j_b$.  If we define the average drift 
velocities carried by the outer-left-edge current, inner-left-edge 
current, inner-right-edge and outer-right-edge currents as $-v_{\rm 
out}^l$, $v_{\rm in}^l$, $-v_{\rm in}^r$ and $v_{\rm out}^r$, 
respectively, with corresponding average electron densities  
$n_{\rm out}^l$, $n_{\rm in}^l$ $n_{\rm in}^r$ and $n_{\rm out}^r$, 
the total current density is  

\begin{equation}
I_i \sim n_{\rm out}^r v_{\rm out}^r -n_{\rm out}^l v_{\rm out}^l
      + n_{\rm in}^l v_{\rm in}^l - n_{\rm in}^r v_{\rm in}^r
\end{equation}
When no current is driven ($I_{\rm drive} = 0$), we have $n_{\rm 
out}^r = n_{\rm out}^l$, $n_{\rm in}^r = n_{\rm in}^l$, $v_{\rm out}^r 
= v_{\rm out}^l$, $v_{\rm in}^r = v_{\rm in}^l$, or in other words the 
current density on the left edge balances out that on the right edge 
and thus $I_i = 0$.  When $I_{\rm drive}$ is switched on ($\delta 
k\neq 0$), $E_{nk}^b$ and $g_n^b (k)$ change.  This affects the drag 
electrons in two ways.  Firstly, on the average there are more 
electrons at $x > 0$ than at $x < 0$ due to $\phi^b (x)$, i.e.  
$n_{\rm in}^r + n_{\rm out}^r > n_{\rm in}^l + n_{\rm out}^l$ and this 
leads to an imbalance of $j_b$.  Secondly, the inner-edge zone, shown 
in Fig.  2, expands into the bulk zone due to the change in $g_n^b 
(k)$.  The resulting total $I_{\rm drag}$ is parallel to $I_{\rm 
drive}$ for $\nu = 0.5$.  As $\nu$ increases, the inner-edge zone 
expands further, due to the increase of $g_n^b (k)$ in the bulk zone 
(the guiding-center coordinate is proportional to $k$), as contrasting 
Fig. 2 with Fig. 5 shows; There is an increase in both $n_{\rm 
in}^l$ and $n_{\rm in}^r$ with $n_{\rm in}^l < n_{\rm in}^r$ and thus 
decreases $I_{\rm drag}$.  
for $\nu$ integer.

As $\nu$ approaches an integer, $g_n^b (k) \to 1$ in the bulk is the 
same as that in the edge zones and there is no bulk activity 
contributing to the drag current.  However, the eigenvalues 
$E_{nk}^b$, shown in Fig. 7, with $n = n_{\rm max}$ near 
$kl_c = 9$ and $n = n_{\rm max} + 1$ near $k l_c = 7$, are degenerate.  
Now the electrons prefer to occupy the inner-edge zone, lower-energy 
sites, with $n_{\rm max} +1$ and $4.8 < |k_2 \ell_c| < 7.6$, rather 
than the outer-edge zone, higher-energy sites, with $n_{\rm max}$ and 
$|k_1 \ell_c| > 8.8$.  This results in the current density shown in 
Fig. 6.  Since the drag current is the total induced current minus
the classical $<{\bf E}>\times {\bf B}$-drift, we  view  these empty states 
as holes  in the $n_{\rm max}$th LL  responsible for $I_{\rm drag}$. These
holes are responsible for the outer-edge currents.  
Since there are more holes at the left edge than at the right edge and 
holes are moving opposite to the outer-edge electrons of the drive-layer, 
the induced current flows opposite to that due to electrons.  
On the other hand, 
electrons in the inner-edge zones of the $n_{\rm max} +1$ LL have a 
total current in the same  direction as that due to  the holes..  
When the sum of these 
two currents exceeds that due to electrons in the bulk  $n_{\rm max}$th LL, 
the total current $I_{\rm drag}$, 
calculated by integrating $j_b (x)$ over the Hall bar, 
flows opposite to $I_{\rm drive}$ in contrast with the $B=0$ case 
where it flows in the direction of $I_{\rm drive}$ for two electron 
layers.  Accordingly, the drag voltage, shown on the right axis of 
Fig.  5,  changes sign when $\nu$ approaches 1 and has a minimum in 
the vicinity of $\nu = 1$.  This change of sign occurs also in $I_{\rm 
drag}$, when allowed to flow, as Eq.  (12) shows.  

The predicted {\it negative} Coulomb drag can
best be tested in an open-circuit configuration, i.e., by measuring 
the voltage difference along the bar to avoid the effect of 
scattering in the drag layer.
A transient measurement should be
conducted to resolve the classical $<{\bf E}>\times {\bf B}$-drift
induced voltage from the momentum-transfer induced $\Delta V_{\rm drag}$
since the response time of the classical drift
is far shorter than that of $\Delta V_{\rm drag}$.  An overshoot
of the measured voltage would signal the {\it negative} Coulomb drag.

Finally, we notice that the induced drag,
for $\nu$ integer or half-integer, is about {\it two-to-three 
orders of magnitude larger}  than 
that at zero field $B$.  This is due to the fact the $B=0$ states 
condense into LLs when $B\neq 0$ and  agrees with the results of Ref. 
\cite{fin}.

\section{CONCLUDING REMARKS}

We have shown, within a self-consistent Hartree
approximation, that the current density in 
Coulomb-coupled {\it  narrow} Hall bars has different current zones that
change with filling factor $\nu$.  In addition to the zero-field
momentum-transfer  current $I_{\rm drag}$, the Hall voltage developed in
the drive layer gives rise to a classical $<{\bf E}> \times {\bf B}$-drift
current $I_{\rm es}$ when the circuit is closed.   As $\nu$ increases from
0.5 towards  1,  $I_{\rm drag}$ or $\Delta V_{\rm drag}$ decreases and,
when $\nu\to 1$, it changes sign. This change occurs because electrons make
energetically favorable {\it  near-edge inter-LL transitions}. We 
expect that the values of $\nu$ where  the drag changes sign   
depend only weakly on 
the initial confining potential. In the present study we took the 
latter square-well in form and evaluated the resulting potential 
self-consistently. Another choice would be an initially parabolic  potential, 
but the results would be qualitatively the same.

It is evident that the above negative drag is neither due to a thermal 
gradient~\cite{bor} nor due to the conventional electron-hole coupling 
since we are dealing with electron layers.  Also, since we have 
neglected tunneling between the layers, it cannot be identified with 
the observed negative drag of Ref.~\cite{pat}.  As we suggested in 
Sec.  IV, it could be tested with time-dependent measurements.

Finally, it should be noted that since we considered an applied 
current in the drive layer in the presence of scattering, i.e., a 
clearly {\it nonequilibrium} case, the resulting drag is a {\it 
dissipative} one and not the {\it non-dissipative} {\it equilibrium} drag 
of Ref.~\cite{vm}. It is a result of the broadening of the
Landau levels introduced by  scattering   coupled with
the nonequilibrium effect as is evident from Eq. ($\ref{degen}$) in which
the occupation is determined self-consistently.  The scattering  
is embodied in $g_n (k)$.  If we were to treat this system as
a non-dissipative one, $g_n (k)$ would be 1 and thus $I_{\rm drag} = 0$.
As for the use of $\delta k$ in the Fermi 
function, to account for the nonequilibrium distribution function, it 
should be said that it was made, in the spirit of Ref.~\cite{wexler}, 
in order to simplify the heavily involved self-consistent 
calculations.  A thorough treatment would have to determine the 
distribution function from the Boltzmann equation.

\acknowledgements

We would like to thank Dr. C. Dharma-wardana and Dr. H. Rubel for fruitful
discussions. This work was supported by the NSERC grants
Nos. OGP0003155 (H. C.  W.T., D. J. W. G.) and OGP0121756 (P. V.).

\ \\
\begin{figure}
\caption{  Schematics of a quantum Hall bar, of length $L$ and of width $W$, 
in a perpendicular magnetic field $B$. The
currents $I_{\rm drag}$ and $I_{\rm int}$ are explained in the text} 
\label{fig1}
\ \\
\caption{  Equilibrium (solid curves) and nonequilibrium (dotted
curves)
densities and current densities as a function of $x$ for $n_{\rm max}
= 1$, $\nu = 0.5$ with $j_0 \equiv e\omega_c/2\pi \ell_c$ and
$\rho_0 \equiv 1/2\pi \ell_c^2$.
The dashed curve is the effective potential $V_{\rm eff}$.} 
 \label{fig2}
 \ \\
\caption{ Nonquilibrium current density as a function of $x$ for
$n_{\rm max} = 1$, $\nu = 0.5$. The solid curve refers to the
left-bottom scales and the dashed curve to the right-top scales.} 
 \label{fig3}
 \ \\
\caption{  Equilibrium LL energies and the weight function as a function
of $k$  for $n_{\rm max}= 1$, $\nu=0.5$.
The dashed line is the chemical potential. The  dots correspond
to $g_1^0 (k)$ and the crosses  to $g_1^b (k)$.} 
\label{fig4}
\ \\ 
\caption{  Equilibrium (solid curves) and nonequilibrium (dotted
curves) densities and current densities, as a function of $x$, and
nonequilibrium weight function (crosses)  as a function of $k$
for $n_{\rm max} = 1$, $\nu = 0.8$.} 
\label{fig5}
\ \\ 
\caption{ Left-bottom scales: Equilibrium (solid curves) and 
nonequilibrium (dotted curves) current densities as a function of
$x$ for $n_{\rm max} = 1$, $\nu = 1$. Right-top scales: The drag
voltage $\Delta V_{\rm drag}$ (crosses) as a function of $\nu$ with
$n_{\rm max} = 1$ and $I_{\rm drive} \simeq 0.08\mu A$.} 
\label{fig6}
\ \\ 
\caption{  Nonequilibrium LL energies as a function of $k$ for
$n_{\rm max} = 1$, $\nu = 1$.  The dashed line is the chemical potential.}
\label{fig7}\end{figure}
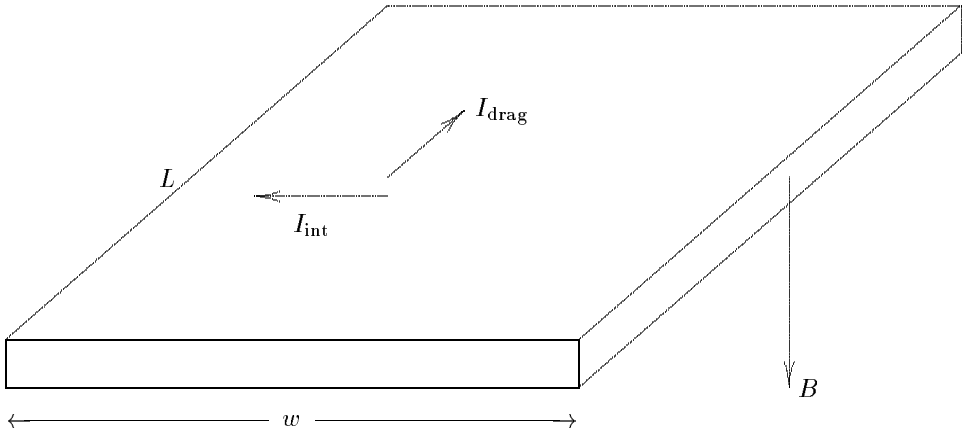
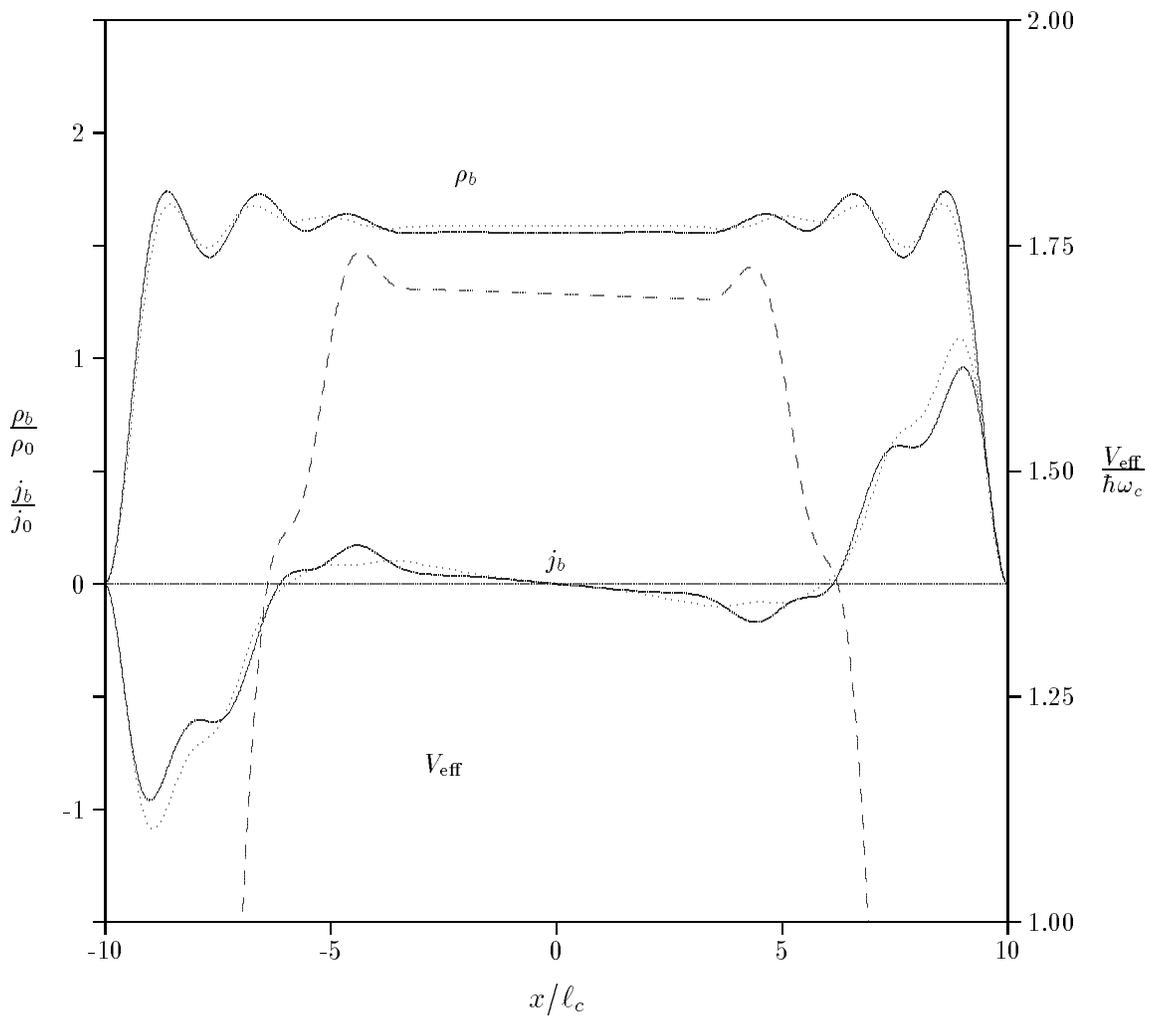
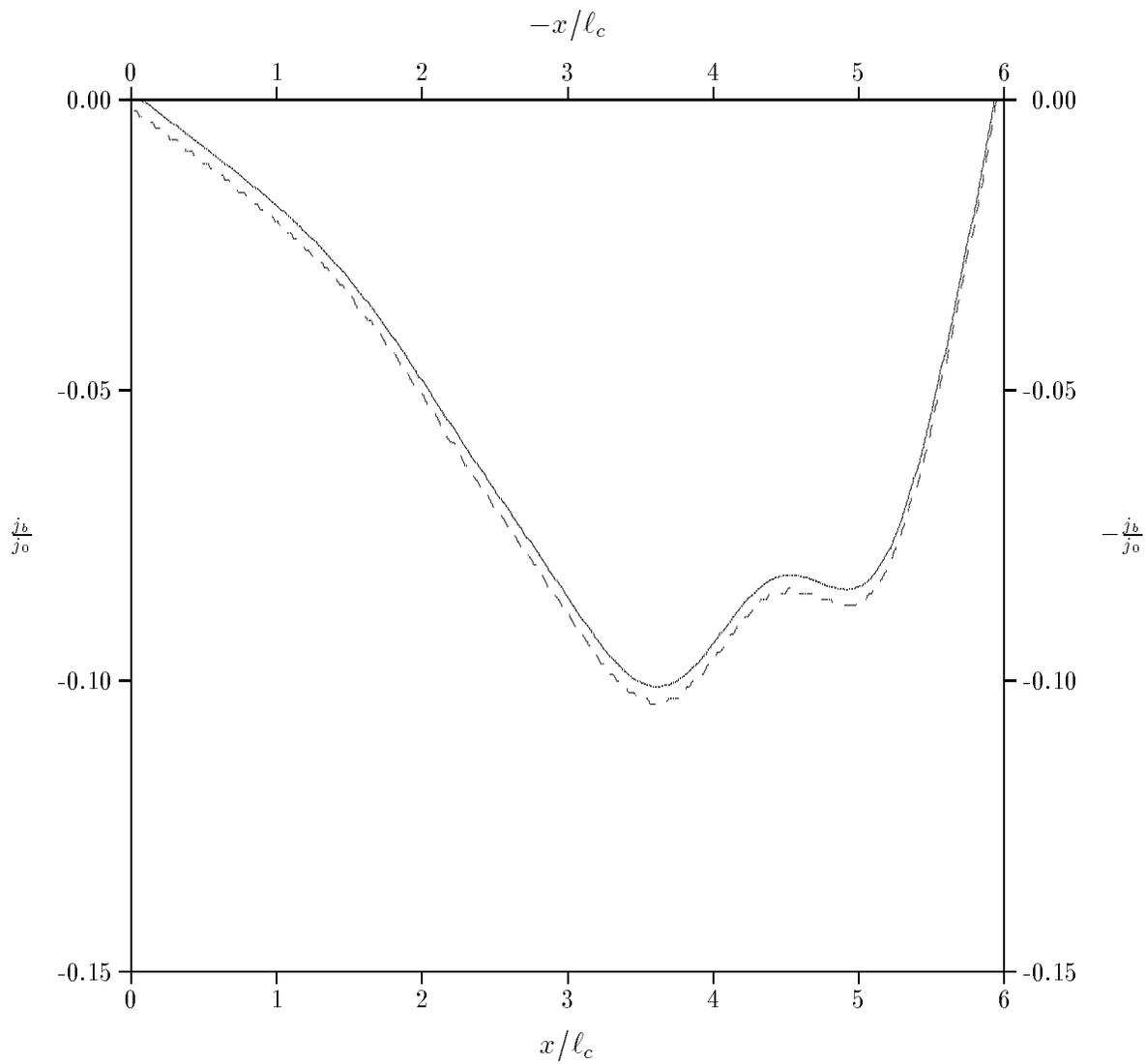
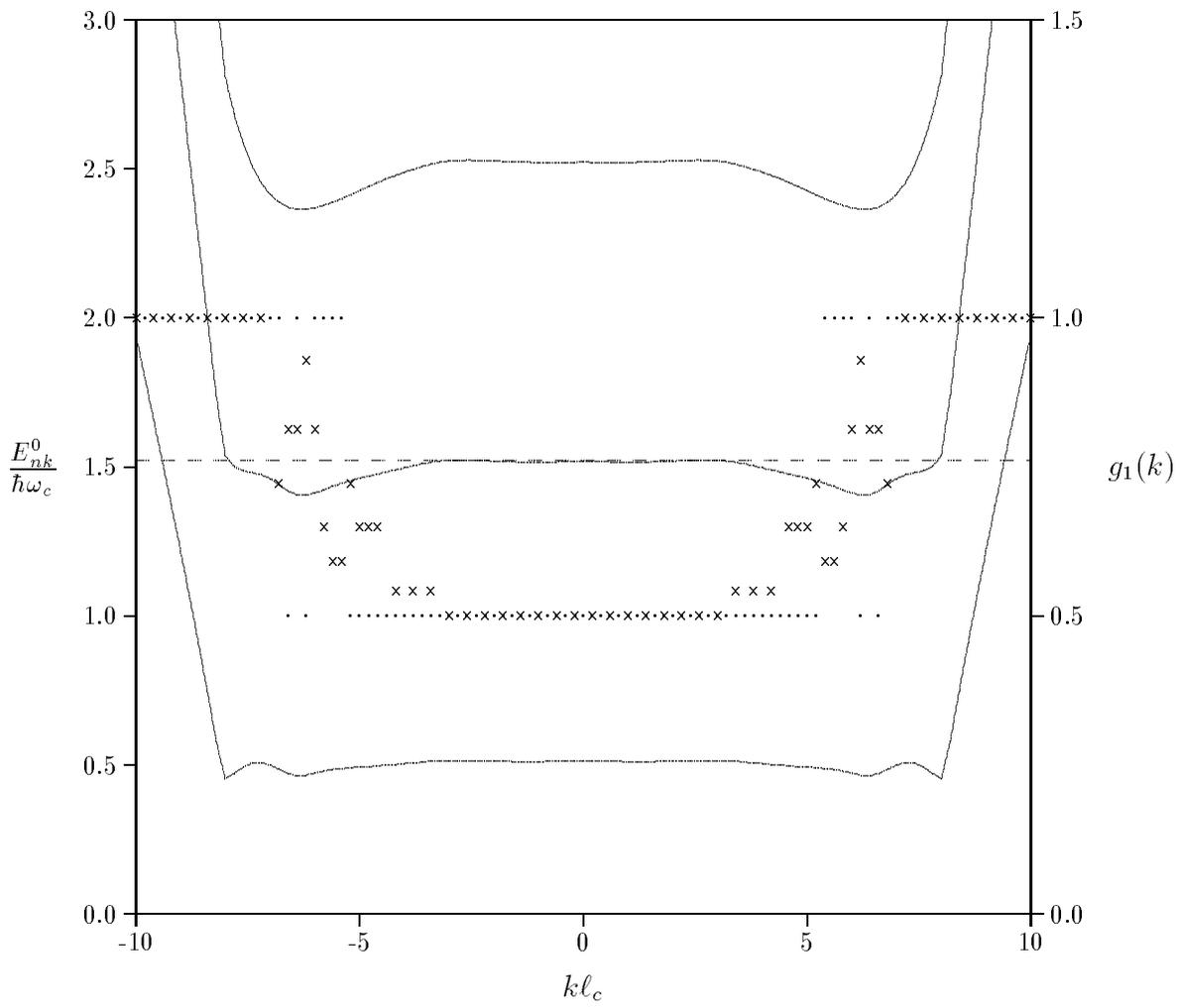
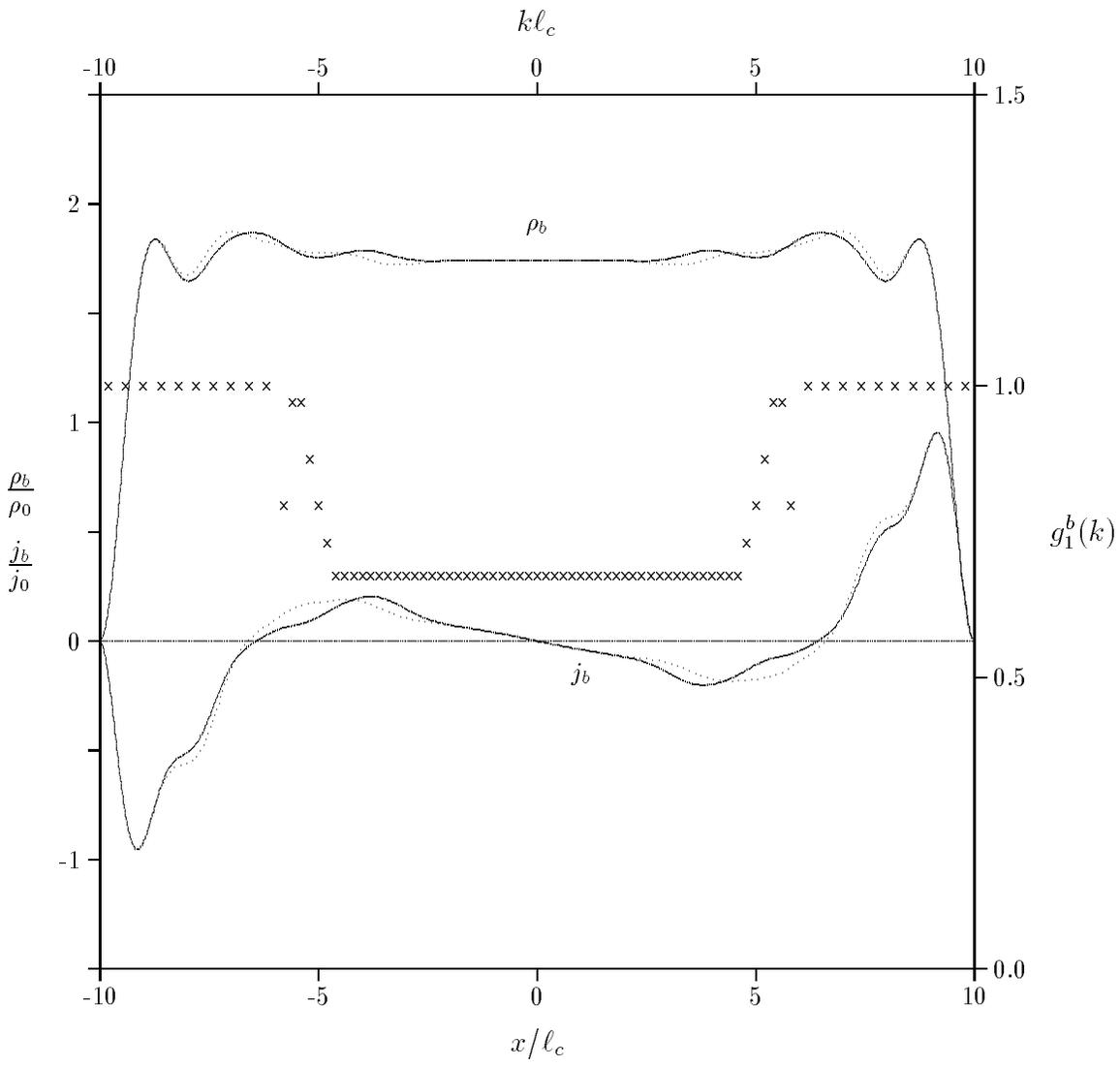
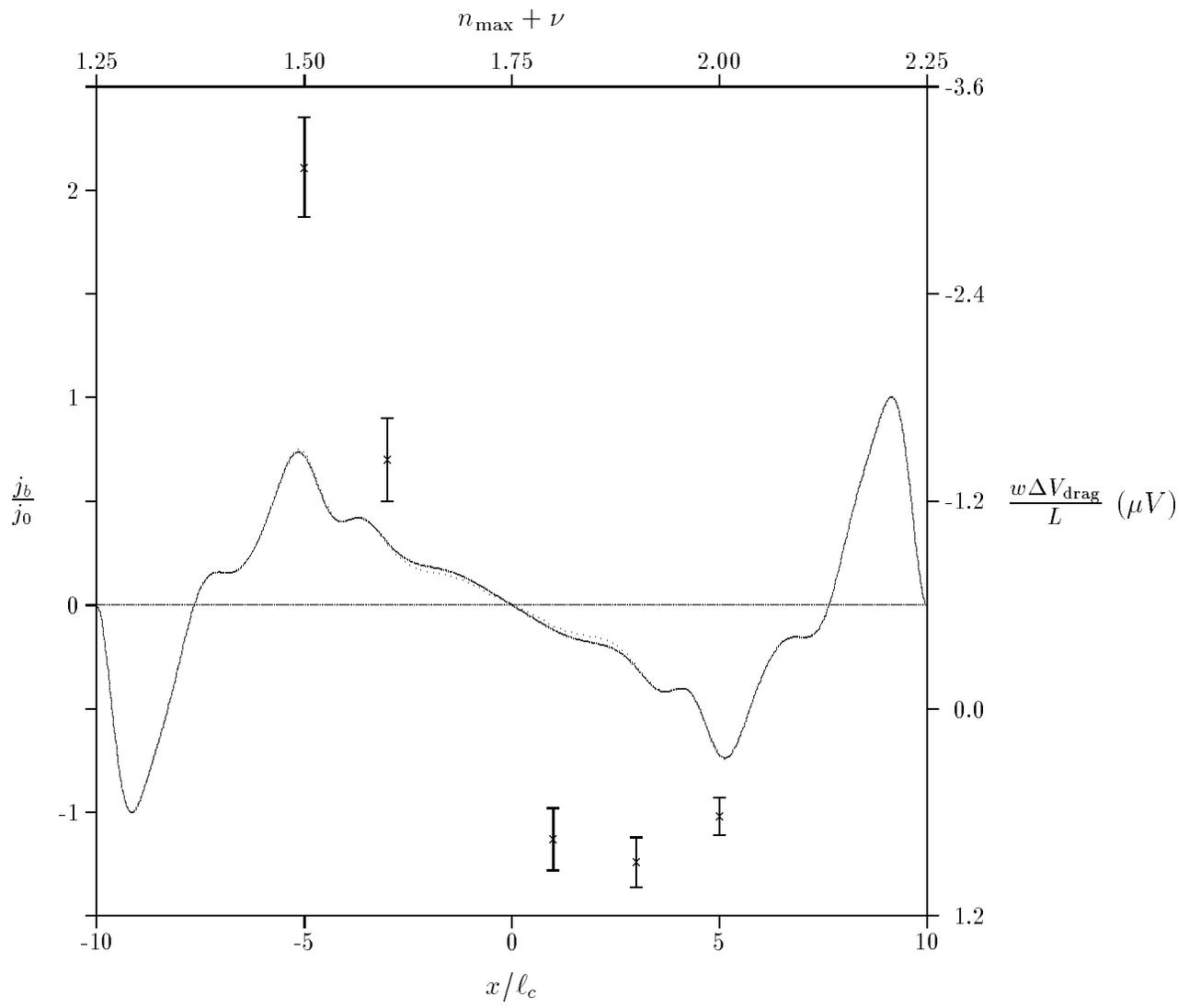
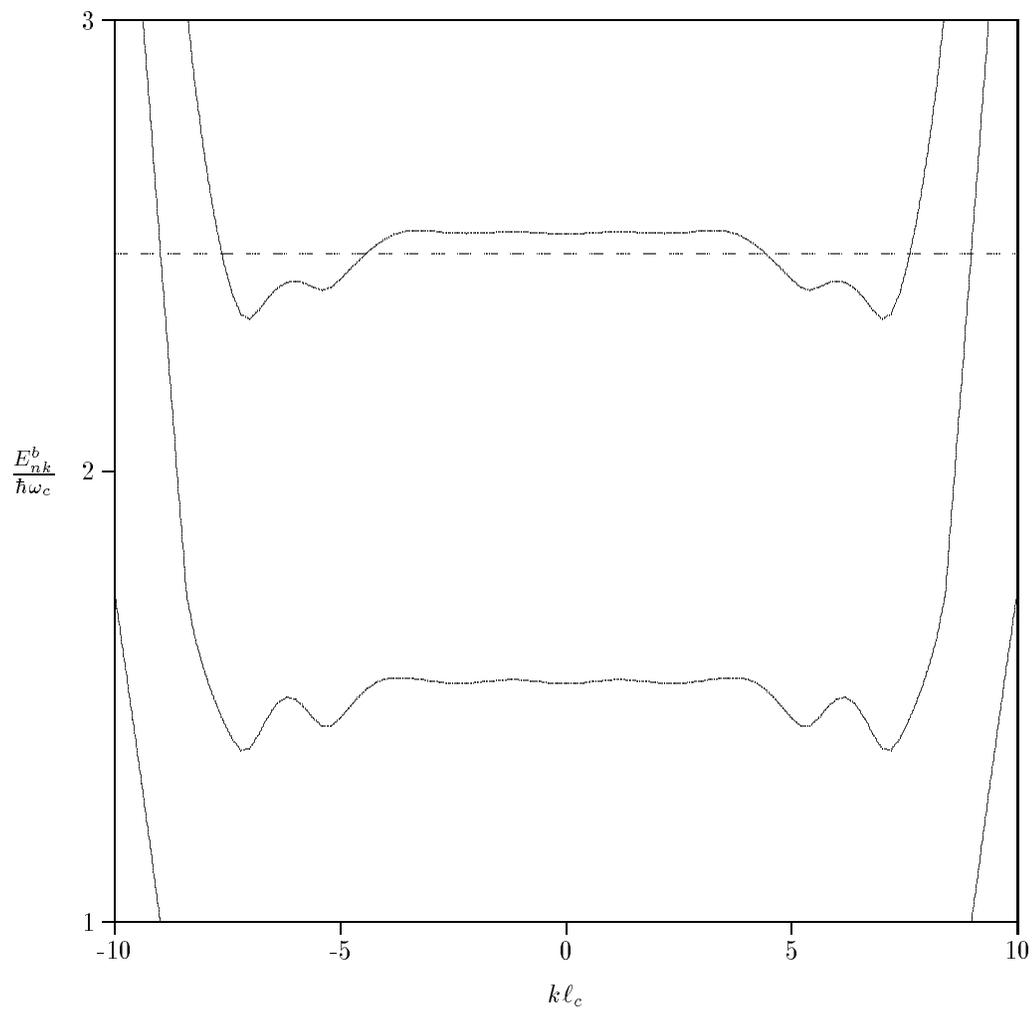
\end{document}